\pdfoutput=1
\documentclass[12pt]{article}

\font\grande=cmr9.5 scaled \magstep4
\font\medio=cmr9.5 scaled \magstep2
\outer\def\beginsection#1\par{\medbreak\bigskip
      \message{#1}\leftline{\bf#1}\nobreak\medskip
\vskip-\parskip
      \noindent}
\usepackage{graphicx} 
\usepackage{mathrsfs}
\begin{document}

\bibliographystyle{unsrt}

\titlepage

\vspace{1cm}
\begin{center}
{\grande Spurious gauge-invariance of higher-order contributions }\\
\vspace{0.5cm}
{\grande to the spectral energy density of the relic gravitons}\\ 
\vspace{1cm}
Massimo Giovannini \footnote{e-mail address: massimo.giovannini@cern.ch}\\
\vspace{1cm}
{{\sl Department of Physics, CERN, 1211 Geneva 23, Switzerland }}\\
\vspace{0.5cm}
{{\sl INFN, Section of Milan-Bicocca, 20126 Milan, Italy}}
\vspace*{1cm}
\end{center}
\vskip 0.3cm
\centerline{\medio  Abstract}
\vskip 0.1cm
In the same way as the energy density associated with the tensor modes of the geometry modifies the evolution of the curvature perturbations, the scalar modes may also indirectly affect the cosmic backgrounds of relic gravitons by inducing higher-order corrections that are only superficially gauge-invariant. This spurious gauge-invariance gets manifest when the effective anisotropic stresses, computed in different coordinate systems, are preliminarily expressed in a form that only depends on the curvature inhomogeneities defined on comoving orthogonal hypersurfaces and on their corresponding time derivatives. Using this observation we demonstrate in general terms that the higher-order contributions derived in diverse coordinate systems coincide when the wavelengths are smaller than the sound horizon defining the evolution of the curvature inhomogeneities but they lead to sharply different results in the opposite limit.  A similar drawback arises when the energy density of the relic gravitons is derived from competing energy-momentum pseudo-tensors and should be consistently taken into account in the related phenomenological discussions.
\noindent
\vspace{5mm}
\vfill
\newpage
Since the energy and the momentum of the gravitational field cannot be localized  \cite{ANIS1} various proposals for the energy density of the relic gravitational waves have been considered through the years \cite{ANIS3,ANIS4,ANIS5,ANIS6a}.  To lowest order the ambiguity of the competing definitions is solved by imposing a number of physical requirements (e.g. the positivity of the energy density both inside and outside the Hubble radius). While the different strategies are all (superficially) gauge-invariant, a careful comparison \cite{ANIS7} implies that the most sound prescription follows from the direct variation of the second-order action with respect to the background fields, as already pointed out long ago by Ford and Parker \cite{ANIS8} (see also Ref. \cite{ANIS8a}).  The spurious gauge-invariance of the energy-momentum pseudo-tensors is ultimately caused by the equivalence principle and necessarily appears in the higher-order processes. For instance the long-wavelength gravitons induce curvature inhomogeneities both during inflation and in the subsequent radiation-dominated phase \cite{ANIS10}. Similarly curvature inhomogeneities may cause higher-order corrections to the stochastic backgrounds of relic gravitons and this second effect involves an effective anisotropic stress which is customarily assessed within the Landau-Lifshitz prescription \cite{ANIS3,ANIS7,ANIS10b, ANIS11}. For the same reasons given above the Landau-Lifshitz approach applied to the scalar modes of the geometry is not gauge-invariant: different and sometimes contradictory statements exist in the current  literature \cite{ANIS12,ANIS13,ANIS14a,ANIS14}. Part of the problem is that the gauge-dependent derivations follow the dynamical evolutions of the pivotal variables in the particular coordinate system where the results have been derived.  To avoid this potential ambiguity the gauge-dependent description will be traded here, from the very beginning, for the curvature perturbations on comoving orthogonal hypersurfaces (${\mathcal R}$ in what follows) and for their first derivative with respect to the conformal time coordinate $\tau$ (${\mathcal R}^{\prime}$ in what follows).  Even though  (${\mathcal R}$, ${\mathcal R}^{\prime}$) are gauge-invariant, the effective anisotropic stress assessed  in diverse coordinate systems always has a distinct functional form in terms of (${\mathcal R}$, ${\mathcal R}^{\prime}$): this mismatch demonstrates beyond any doubt that the whole approach is intrinsically {\em not} gauge-invariant.

Consider a perfect, relativistic and irrotational fluid with energy-momentum tensor $T_{\mu}^{\,\,\nu} = (\rho_{t} + p_{t}) u_{\mu} \, u^{\nu} - p_{t} \delta_{\mu}^{\nu}$ where $\rho_{t}$, $p_{t}$ and $u_{\mu}$ denote, respectively, the total energy density, the pressure and the four-velocity. In a conformally flat background geometry\footnote{The conformally flat background metric will therefore be of the type $\overline{g}_{\mu\nu}(\tau) = a^2(\tau) \, \eta_{\mu\nu}$ where $\eta_{\mu\nu}$ defines the Minkowski metric with signature $(+,\,-,\,-,\,-)$; $a(\tau)$ is the scale factor and $\tau$ denotes the conformal time coordinate.}, the gauge-invariant curvature inhomogeneities corresponding to the normal modes of the gravitating fluid evolve, in Fourier space, as \cite{ANIS14b,ANIS14d}: 
\begin{equation}
{\mathcal R}_{\vec{k}}^{\prime\prime} + 2 \frac{z_{t}^{\prime}}{z_{t}} {\mathcal R}_{\vec{k}}^{\prime} + k^2 \, c_{st}^2 {\mathcal R}_{\vec{k}} =0, \qquad z_{t} = \frac{a^2 \sqrt{ p_{t} + \rho_{t}}}{{\mathcal H} c_{st}},\qquad c_{st}^2 = \frac{\partial_{\tau} p_{t}}{\partial_{\tau} \rho_{t}}= \frac{p_{t}^{\prime}}{\rho_{t}^{\prime}},
\label{ONEa}
\end{equation}
where the prime denotes a derivation with respect to the conformal time coordinate $\tau$;
${\mathcal H} = a\, H$ and $H$ is the Hubble expansion rate;  $c_{st}$ is the total sound speed of the plasma.  The solution of Eq. (\ref{ONEa}) in the short-wavelength limit (i.e. $k^2 c_{s}^2 \gg |z_{t}^{\prime\prime}/z_{t}|$) follows from Wentzel-Kramers-Brillouin (WKB) approximation:
\begin{eqnarray}
{\mathcal R}_{\vec{k}}(\tau) &=& \frac{C_{\vec{k}}}{z_{t}\,\sqrt{2 \,k \, c_{st}}} \,\cos{[ k\, r_{s}(\tau)]} + \frac{D_{\vec{k}}}{z_{t}\,\sqrt{2 \,k \, c_{st}}}\,\sin{[ k\, r_{s}(\tau)]}, 
\nonumber\\
r_{s}(\tau) &=& \int_{\tau_{i}}^{\tau} \, c_{st}(\tau) \, d\tau,\qquad \frac{c_{st}^{\prime}}{2 c_{st}} < k \, c_{st},
\label{ONEb}
\end{eqnarray}
where $C_{\vec{k}}$ and $D_{\vec{k}}$ are two constants (possibly determined from the boundary conditions) and $r_{s}(\tau)$ defines the sound horizon. For short, the wavelengths satisfying Eq. (\ref{ONEb}) will be said to be inside the sound horizon (i.e. $ k\,r_{s}(\tau) \gg 1$). In the opposite limit (i.e. $k\,r_{s}(\tau) \ll 1$) the  wavelengths are larger than the sound horizon and  the solution of Eq. (\ref{ONEa}) can be determined by iteration from the following integral equation:
\begin{equation}
{\mathcal R}_{\vec{k}}(\tau) = {\mathcal R}_{\vec{k}}(\tau_{ex}) + {\mathcal R}_{\vec{k}}^{\,\prime}(\tau_{ex})\, \int_{\tau_{ex}}^{\tau} \frac{z_{ex}^2}{z_{t}^2(\tau_{1})} \, d\tau_{1} - k^2 \int_{\tau_{ex}}^{\tau} \frac{d\tau_{2}}{z_{t}^2(\tau_{2})}\, \int_{\tau_{ex}}^{\tau_{2}} c_{st}^2(\tau_{1})\, z_{t}^2(\tau_{1})\, {\mathcal R}_{\vec{k}}(\tau_{1}) \, d\tau_{1}.
\label{FOURa}
\end{equation}
In the preceding expression $\tau_{ex}$ denotes the time at which the given scale exits the Hubble radius (i.e. $k c_{st} \tau_{ex} \simeq 1$); during inflation $z_{t} \to z_{\varphi} = a\, \varphi^{\prime}/{\mathcal H}$ (where $\varphi$ is the inflaton) so that $k \tau_{ex} \simeq 1$ and the sound horizon coincides, in practice, with the Hubble radius. 

The second-order scalar fluctuations of the Einstein tensor and of the matter sources determine the effective anisotropic stress in a specific coordinate system. Following the standard Landau-Lifshitz strategy \cite{ANIS3,ANIS7,ANIS10b,ANIS11}  the first-order scalar equations in that particular gauge are then used to simplify the obtained expressions. Finally the total spatial derivatives\footnote{Terms of the form $\partial_{i} ( f\, \partial_{j} g) $ (where $f$ and $g$ are two generic first-order fluctuations) do not contribute to the effective anisotropic stress since their projection over the two tensor polarizations appearing in Eq. (\ref{SIXa}) vanishes.} do not contribute to the effective anisotropic stress.  If the Einstein's equations are written in the form ${\mathcal G}_{\mu}^{\,\,\,\nu} = \ell_{P}^2 \, T_{\mu}^{\,\,\,\nu}$ (where ${\mathcal G}_{\mu\nu}$ denotes the Einstein tensor and 
$\ell_{P} = \sqrt{8 \pi G}$)  the evolution of the tensor modes of the geometry formally follows from
\begin{equation}
\delta_{t}^{(1)} {\mathcal G}_{i}^{\,\,\,j} = - \delta_{s}^{(2)} {\mathcal G}_{i}^{\,\,\,j} + \ell_{P}^2 \delta_{s}^{(2)}\, T_{i}^{\,\,\,j}\qquad \Rightarrow \Pi_{i}^{(X)\,\,j}(\vec{x}, \tau) = \ell_{P}^2 \, \delta^{(2)}_{s} T_{i}^{\,\,\,j} - \delta_{s}^{(2)} {\mathcal G}_{i}^{\,\,\,j},
 \label{FIVE}
\end{equation}
where $\delta_{t}^{(1)}$ denotes the first-order (tensor) fluctuation while 
$\delta_{s}^{(2)}$ is the second-order (scalar) fluctuation of the corresponding quantity; 
the superscript $X$ reminds that the scalar modes are defined in the coordinate system $X$. The first-order scalar inhomogeneities evolve then according to $\delta_{s}^{(1)} {\mathcal G}_{\mu}^{\nu} = \ell_{P}^2 \delta_{s}^{(1)} T_{\mu}^{\nu}$ while the tensor amplitudes and the corresponding anisotropic stress carry two tensor polarizations:
\begin{equation}
h_{i}^{\,\,\, j} (\vec{q}, \tau) =\sum_{\lambda= \oplus,\, \otimes} \, e^{(\lambda)\,\,\,j}_{i}(\hat{q}) \, \, h_{\lambda}(\vec{q},\tau), \qquad \Pi_{i}^{(X)\,\,\,j}(\vec{q},\tau) = \sum_{\lambda= \oplus,\, \otimes}  \, e^{(\lambda)\,\,\,j}_{i}(\hat{q}) \, \Pi_{\lambda}^{(X)}(\vec{q},\tau).
\label{SIXa} 
\end{equation}
For a gravitational wave propagating in the $\hat{q}$ direction the polarizations are defined as $e_{i\,j}^{(\oplus)}(\hat{q}) = \hat{m}_{i} \,\hat{m}_{j} - \hat{n}_{i} \,\hat{n}_{j}$ and $e_{i\,j}^{(\otimes)}(\hat{q}) = \hat{m}_{i} \,\hat{n}_{j} + \hat{n}_{i} \,\hat{m}_{j}$ 
where $\hat{m}$, $\hat{n}$ and $\hat{q}$ are three mutually orthogonal unit vectors. 
From Eqs. (\ref{FIVE}) and (\ref{SIXa}) each tensor polarization obeys the following evolution 
equation: 
\begin{equation}
h_{\lambda}^{\prime\prime} + 2 {\mathcal H} h_{\lambda}^{\prime} + q^2 h_{\lambda} = - 2 \,\ell_{P}^2 \,a^2(\tau) \, \Pi_{\lambda}^{(X)}.
\label{SIX}
\end{equation}
The solution of Eq. (\ref{SIX}) for $h_{\lambda}$ and $\partial_{\tau} h_{\lambda}$ is formally expressed in terms 
of the corresponding Green's functions $G[ q ( \xi- \tau)]$ and $\widetilde{\,G\,}[q ( \xi -\tau)]$:
\begin{eqnarray}
h_{\lambda}^{(X)}(\vec{q},\tau) &=& \overline{h}_{\lambda}(\vec{q},\tau) - 2 \ell_{P}^2 \int_{\tau_{i}}^{\tau} d \xi \, a^2(\xi) \,G[ q ( \xi- \tau)] \, 
\Pi_{\lambda}^{(X)}(\vec{q}, \xi), 
\nonumber\\
H_{\lambda}^{(X)}(\vec{q},\tau) &=& \overline{H}_{\lambda}(\vec{q},\tau) - 2 \ell_{P}^2  \int_{\tau_{i}}^{\tau} d \xi \, a^2(\xi) \,\widetilde{\,G\,}[ q ( \xi- \tau)] \, 
\Pi_{\lambda}^{(X)}(\vec{q}, \xi), 
\label{SEVEN}
\end{eqnarray}
where $H^{(X)}_{\lambda} = \partial_{\tau} h^{(X)}_{\lambda}$ and $\overline{H}_{\lambda} =\partial_{\tau}\overline{h}_{\lambda} $; the overline distinguishes the (gauge-invariant) first-order contributions  from their second-order (gauge-dependent) counterparts.

In the $X$-gauge the effective anisotropic stress will depend on the pivotal variables 
of that specific coordinate system. For instance, in the conformally Newtonian gauge, only the longitudinal entries of the metric are perturbed $\delta_{s}^{(1)} \, g_{00}= 2 \,a^2 \, \phi$ and $\delta^{(1)} g_{ij} = 2 \, a^2\, \psi\, \delta_{ij}$; hence the aforementioned gauge will be referred to as the $L$-gauge  and using the definition of Eq. (\ref{FIVE}), a straightforward calculation leads to the following form of the effective anisotropic stress:
\begin{eqnarray}
\Pi_{i}^{(L)\,\,j}(\vec{x},\tau) &=& \frac{1}{ \ell_{P}^2 \, a^2} \biggl[ \partial_{i} \psi \partial^{j} \psi - \partial_{i} \phi \partial^{j} \phi  - \partial_{i} \phi \partial^{j} \psi  
- \partial_{i} \psi \partial^{j} \phi 
\nonumber\\
&+& 2 \psi \partial_{i} \partial^{j} ( \phi - \psi) -\frac{2 ({\mathcal H}^2 - {\mathcal H}^{\prime})}{{\mathcal H}^2} \partial_{i}(\psi^{\prime } + {\mathcal H} \phi) \partial^{j} (\psi^{\prime } + {\mathcal H} \phi)
\biggr].
\label{EIGHT}
\end{eqnarray}
Equation (\ref{EIGHT}) only depends, as anticipated, on the pivotal variables of the $L$-gauge. 
The longitudinal entries of the perturbed metric can be however traded
for  ${\mathcal R}_{\vec{k}}$ and ${\mathcal R}_{\vec{k}}^{\prime}$ obeying Eq. (\ref{ONEa}):
\begin{equation}
{\mathcal R}_{\vec{k}} + \psi_{\vec{k}} = - \frac{{\mathcal H}}{{\mathcal H}^2 - {\mathcal H}^{\prime}} ( {\mathcal H} \phi_{\vec{k}} + 
\psi_{\vec{k}}^{\prime} ), \qquad {\mathcal R}_{\vec{k}}^{\prime} = \frac{2 \, a^2 \, k^2 \, \psi_{\vec{k}}}{ \ell_{P}^2 \, {\mathcal H}\, z_{t}^2}.
\label{NINE}
\end{equation}
 Inserting Eq. (\ref{NINE}) into Eq. (\ref{EIGHT}) the effective anisotropic stress becomes
\begin{eqnarray}
\Pi_{i\, j}^{(L)}(\vec{q}, \tau) &=& - \frac{2 ({\mathcal H}^2 - {\mathcal H}^{\prime})}{(2\pi)^{3/2}\,\ell_{P}^2\, a^2(\tau)\,{\mathcal H}^2 } \int\, d^{3}k\, \, k_{i} \,\, k_{j} \biggl\{ {\mathcal R}_{\vec{k}} \, {\mathcal R}_{\vec{q} - \vec{k}} + \frac{{\mathcal H}^2 - {\mathcal H}^{\prime}}{{\mathcal H}} \biggl[ \frac{{\mathcal R}_{\vec{k}} \, {\mathcal R}_{\vec{q} - \vec{k}}^{\prime}}{c_{st}^2  \, |\vec{q} - \vec{k}|^2} 
\nonumber\\
&+& \frac{{\mathcal R}_{\vec{k}}^{\prime} \, {\mathcal R}_{\vec{q} - \vec{k}} }{c_{st}^2  \, k^2}\biggr] + 
\frac{( 2 {\mathcal H}^2 - {\mathcal H}^{\prime}) ({\mathcal H}^2 - {\mathcal H}^{\prime})}{{\mathcal H}^2 \, c_{st}^4 \, k^2 \, |\vec{q}- \vec{k}|^2 } \, {\mathcal R}_{\vec{k}}^{\prime}\,  {\mathcal R}_{\vec{q} - \vec{k}}^{\prime}\biggr\}.
\label{TEN}
\end{eqnarray}
The neutrinos free-stream after electron-positron annihilation and their (tensor) anisotropic stress suppresses the spectral energy density of the 
relic gravitons of 10 \% \cite{ANIS15a}. The (scalar) anisotropic stress $\pi_{\nu}$ determines the difference between the two longitudinal fluctuations of the metric as $k^2 (\phi_{\vec{k}} - \psi_{\vec{k}}) = - 3 \ell_{P}^2 \pi_{\nu}/2 \to 0$. This correction can be  neglected  for the present purposes since it is of even higher-order. We shall therefore posit 
that $\phi_{\vec{k}} \simeq \psi_{\vec{k}}$, as customarily assumed in the discussions of these effects.   

The effective anisotropic stress changes in a different coordinate system since the pivotal variables on the new gauge will be necessarily different from the old ones. The anisotropic stress computed in the new coordinate system will always be expressible in terms of ${\mathcal R}_{\vec{k}}$ and ${\mathcal R}_{\vec{k}}^{\prime}$ so that the results computed in competing gauges will not coincide.  For instance in the coordinate system where the spatial curvature is uniform the perturbed entries of the metric are 
$ \delta_{s}^{(1)}\, g_{00} = 2 \, a^2\, \phi$  and $\delta^{(1)}_{s} g_{0i} = - a^2 \partial_{i} B$. Within this notation (which is incidentally the one adopted in Ref. \cite{ANIS14d}) the effective anisotropic stress of Eq. (\ref{FIVE}) becomes, in Fourier space,
\begin{equation}
\Pi_{ij}^{(U)}(\vec{q}, \tau) = - \frac{1}{(2\pi)^{3/2} \, \ell_{P}^2\, a^2(\tau)} \int d^{3} k \,k_{i}\,k_{j} \biggl[  \biggl( \frac{{\mathcal H}^2 + {\mathcal H}^{\prime}}{{\mathcal H}^2 - {\mathcal H}^{\prime}}\biggr) \phi_{\vec{k}} 
\phi_{\vec{q} - \vec{k}}+ \frac{1}{2} \biggl( \phi_{\vec{k}}^{\prime} \, B_{\vec{q} - \vec{k}} + \phi_{\vec{q} - \vec{k}}^{\prime} \, B_{\vec{k}} \biggr) \biggr],
\label{TWELVE}
\end{equation}
and it is exactly the analog to Eq. (\ref{TEN}) but in the gauge where the spatial curvature is uniform.
In the $U$-gauge the absence of the scalar anisotropic stress implies  $ B_{\vec{k}}^{\prime} + 2 {\mathcal H} B_{\vec{k}} = - \phi_{\vec{k}}$ and the relation of $\phi_{\vec{k}}$ and $B_{\vec{k}}$ to the curvature perturbations 
on comoving orthogonal hypersurfaces is simply given by \cite{ANIS14d}:
\begin{equation}
\phi_{\vec{k}} = - \biggl(\frac{{\mathcal H}^2 - {\mathcal H}^{\prime}}{{\mathcal H}^2} \biggr) \, {\mathcal R}_{\vec{k}}, \qquad 
B_{\vec{k}} = - \frac{\ell_{P}^2 \, z_{t}^2(\tau)}{2\, a^2(\tau)\, k^2}\, {\mathcal R}_{\vec{k}}^{\prime}.
\label{THIRTEEN}
\end{equation}
If  Eq. (\ref{THIRTEEN}) is inserted into Eq. (\ref{TWELVE}) the canonical form of $\Pi_{ij}^{(U)}(\vec{q}, \tau)$ becomes:
\begin{eqnarray}
\Pi_{ij}^{(U)}(\vec{q}, \tau) &=& - \frac{({\mathcal H}^2 - {\mathcal H}^{\prime})^2}{(2\pi)^{3/2} \,\ell_{P}^2\, a^2(\tau)\, {\mathcal H}^4} \int d^{3} k\,\, k_{i}\,k_{j} \,\biggl\{  \biggl( \frac{{\mathcal H}^2 + {\mathcal H}^{\prime}}{{\mathcal H}^2 - {\mathcal H}^{\prime}}\biggr) {\mathcal R}_{\vec{k}} \,{\mathcal R}_{\vec{q} - \vec{k}} 
\nonumber\\
&+& \frac{3}{2} {\mathcal H} (w - c_{st}^2) \biggl[ \frac{{\mathcal R}_{\vec{q} - \vec{k}}^{\prime} \, {\mathcal R}_{\vec{k}}}{|\vec{q} - \vec{k}|^2 \, c_{st}^2} + \frac{{\mathcal R}_{\vec{q} - \vec{k}}\, {\mathcal R}_{\vec{k}}^{\prime}}{k^2 \, c_{st}^2} \biggr] +  \frac{k^2 + |\vec{q} - \vec{k}|^2}{2\,c_{st}^2 \, |\vec{q} - \vec{k}|^2\,k^2} {\mathcal R}_{\vec{k}}^{\prime}  
{\mathcal R}_{\vec{q} -\vec{k}}^{\prime} \biggr\}.
\label{FOURTEEN}
\end{eqnarray}
Since  Eqs. (\ref{TEN}) and (\ref{FOURTEEN}) have a markedly different form (but depend on  the same gauge-invariant variables) the whole expression is not invariant under infinitesimal coordinate transformations and the gauge-invariance is only spurious.

The differences between Eqs. (\ref{TEN}) and (\ref{FOURTEEN}) are not determined by  evolutionary features occurring in competing coordinate systems (as sometimes argued in the past) but rather by the lack of localization of the energy-momentum of the gravitational field ultimately coming from 
 the equivalence principle. In fact the 
same spurious gauge-invariance arises in the derivation of the competing 
energy-momentum (pseudo)tensors of the relic gravitons.  For instance the energy-momentum pseudo-tensor obtained from the variation of the effective 
action of the relic gravitons with respect to the background metric leads to the energy density firstly derived by Ford and Parker
\cite{ANIS8}:
\begin{equation}
\rho_{gw} = \frac{1}{8 \ell_{\mathrm{P}}^2 a^2} \biggl[ \partial_{\tau} h_{k \ell}\, \partial_{\tau}h^{k \ell} + \partial_{m} h_{k\ell} \partial^{m} h^{k\ell}\biggr],
\label{FIFTEEN}
\end{equation}
The energy-momentum pseudo-tensor following instead
from the Landau-Lifshitz strategy \cite{ANIS3,ANIS7,ANIS10b} is computed from the second-order tensor variation  (i.e. ${\mathcal T}_{\mu}^{\,\,\,\nu} = - \delta_{t}^{(2)} \, {\mathcal G}_{\mu}^{\,\,\,\nu}/\ell_{P}^2$ in full analogy with Eq. (\ref{FIVE})); in the latter case the energy density of the relic gravitons is still gauge-invariant (i.e. it only depends on $h_{ij}$ and its first time derivative) but it differs from Eq. (\ref{FIFTEEN}) by a factor ${\mathcal H} \, (\partial_{\tau}h_{k\ell })\, h^{k\ell}/(a^2 \ell_{\rm P}^2)$. 
As consequence the Landau-Lifshitz \cite{ANIS3} and the Ford-Parker \cite{ANIS8} strategies coincide 
for typical frequencies larger than the rate of variation of the geometry (i.e. $ k \gg a\, H$) but  in the opposite limit Eq. (\ref{FIFTEEN}) leads to a positive semi-definite energy density while the Landau-Lifshitz approach violates the weak energy condition \cite{ANIS7,ANIS10b}.

Bearing in mind the analogy to the spurious gauge-invariance of the energy-momentum pseudo-tensors, we shall now demonstrate that Eqs. (\ref{TEN}) and (\ref{FOURTEEN})
coincide for typical wavelengths shorter than the sound horizon of Eq. (\ref{ONEb}) 
but differ outside of it. Without assuming of enforcing any specific evolution of the background we recall, 
from Eq. (\ref{ONEb}), that inside the sound horizon (i.e. up to corrections that are negligible in the 
limit $k \, c_{st} \gg a\, H$),  the curvature perturbations and their derivatives are approximately 
related as ${\mathcal R}_{\vec{k}}^{\prime} \simeq k \,c_{st}\, {\mathcal R}_{\vec{k}}$.  
If Eq. (\ref{TEN}) is expanded in the limit $k \, c_{s t} \gg H \, a$ and $|\vec{q} - \vec{k}| \, c_{st} \gg a\, H$ 
the expression of the effective anisotropic stress becomes\footnote{For short these limit can be dubbed as 
$k \, c_{st} \, \tau \gg 1$, $|\vec{q} - \vec{k}| \, c_{st}\, \tau \gg 1$ with $(k \, c_{st} \, \tau)/(|\vec{q} - \vec{k}| 
\, c_{st}\, \tau ) \to 1$.}:
\begin{eqnarray}
\Pi_{ij}^{\,(L)}(\vec{q}, \tau)  &=& - \frac{2 ({\mathcal H}^2 - {\mathcal H}^{\prime})}{(2\pi)^{3/2}\,\ell_{P}^2\, a^2(\tau)\,{\mathcal H}^2 } \int\, d^{3}k\, \, k_{i} \,\, k_{j} {\mathcal R}_{\vec{k}} \, {\mathcal R}_{\vec{q} - \vec{k}}\, \biggl\{ 1+ ({\mathcal H}^2 - {\mathcal H}^{\prime})  \frac{( k+ |\vec{q} - \vec{k}|)}{c_{st}  \, {\mathcal H}\, k\, |\vec{q} - \vec{k}|}\nonumber\\
&+& \frac{( 2 {\mathcal H}^2 - {\mathcal H}^{\prime})}{{\mathcal H}^2 \, c_{st}^2 \, k \, |\vec{q}- \vec{k}| } + \, .\, .\,. \biggr\},
\label{FIVETEEN}
\end{eqnarray}
where the ellipses stand for the higher-order contributions. 
The first term at the right hand side of Eq. (\ref{FIVETEEN}) 
dominates in the limit $k\, c_{st} \gg H a$  while the two remaining contributions are 
of higher order. The same steps leading to Eq. (\ref{FIVETEEN}) 
can be repeated for the expression of $\Pi_{ij}^{\,(U)}(\vec{q},\tau)$ derived in Eq. (\ref{FOURTEEN}) and the corresponding result will be:
\begin{eqnarray}
\Pi_{i j}^{(U)}(\vec{q},\tau) &=& - \frac{({\mathcal H}^2 - {\mathcal H}^{\prime})^2}{(2\pi)^{3/2} \, \ell_{P}^2\, {\mathcal H}^4 a^2}
\,\, \int d^{3} k \,\, k_{i} \, k_{j} {\mathcal R}_{\vec{k}} \, {\mathcal R}_{\vec{q} - \vec{k}} \biggl\{ 1 + \frac{{\mathcal H}^2 
+ {\mathcal H}^{\prime}}{{\mathcal H}^2 - {\mathcal H}^{\prime}} 
\nonumber\\
&+& \frac{3}{2} {\mathcal H} \frac{( w - c_{st}^2)\,(k + |\vec{q} - \vec{k}|)}{c_{st}\,|\vec{q} - \vec{k}| \, k}  + \,.\,.\,.\biggr\}.
\label{EIGHTEEN}
\end{eqnarray}
Equations (\ref{FIVETEEN}) and (\ref{EIGHTEEN}) demonstrate 
that the leading terms of both expansions are the same. Therefore, as long as 
the corresponding wavelengths are shorter than the sound horizon the two expressions 
of the effective anisotropic stress will coincide up to subleading corrections:
\begin{equation}
\Pi_{ij}^{(L)}(\vec{q}, \tau) = \Pi_{ij}^{(U)}(\vec{q},\tau) + {\mathcal O}\biggl(\frac{a\, H}{k \, c_{st}}\biggr) + {\mathcal O}\biggl(\frac{a\, H}{|\vec{q} - \vec{k}| \, c_{st}}\biggr)  + {\mathcal O}\biggl(\frac{a^2\, H^2}{k \, |\vec{q} - \vec{k}| \, c_{st}^2}\biggr) +\,.\,.\,.
\label{NINETEEN}
\end{equation}
When the typical wavelengths are larger than the sound horizon the evolution 
of the curvature perturbations follows from Eq. (\ref{FOURa}) which will now be inserted into Eqs. (\ref{TEN}) and (\ref{FOURTEEN}); the resulting general expression is:
\begin{equation}
\Pi_{ij}^{(X)}(\vec{q},\tau) = - \frac{1}{(2 \pi)^{3/2} \,a^2(\tau)\, \ell_{P}^2} \int d^{3} k \, k_{i} \, k_{j} \, A^{(X)}(\vec{k}, \vec{q}, \tau) \biggl[ 1  +   {\mathcal O}\biggl(\frac{k \, c_{st}}{a H}\biggr) + 
{\mathcal O}\biggl(\frac{|\vec{q} - \vec{k}| \, c_{st}}{a H}\biggr) + .\,.\,.\biggr],
\label{TWENTYONE}
\end{equation}
where $X= L,\, U$ and the explicit forms of the leading contributions are:
\begin{eqnarray}
A^{(L)}(\vec{k}, \vec{q}, \tau) &=& 2 \frac{(2 {\mathcal H}^2 - {\mathcal H}^{\prime})({\mathcal H}^2 - {\mathcal H}^{\prime})^2}{{\mathcal H}^4\, \, k^2 \, |\vec{q} -\vec{k}|^2 \, c_{st}^4} \biggl(\frac{z_{ex}}{z_{t}}\biggr)^{4} \, {\mathcal R}_{\vec{k}}^{\,\,\prime}(\tau_{ex})\,  {\mathcal R}_{\vec{q} - \vec{k}}^{\,\,\prime}(\tau_{ex}),
\label{TWENTYTWO}\\
A^{(U)}(\vec{k}, \vec{q}, \tau) &=& \frac{({\mathcal H}^2 - {\mathcal H}^{\prime})^2\,[ k^2 + |\vec{q} - \vec{k}|^2 }{2 \, {\mathcal H}^4\, k^2 \, c_{st}^2 |\vec{q} - \vec{k}|^2} \biggl(\frac{z_{ex}}{z_{t}}\biggr)^{4} \, {\mathcal R}_{\vec{k}}^{\,\,\prime}(\tau_{ex})\,  {\mathcal R}_{\vec{q} - \vec{k}}^{\,\,\prime}(\tau_{ex}).
\label{TWENTYTHREE}
\end{eqnarray}
From the ratio between Eqs. (\ref{TWENTYTWO}) and (\ref{TWENTYTHREE}) we see that 
 $A^{(U)}(\vec{k}, \vec{q}, \tau)/A^{(L)}(\vec{k}, \vec{q}, \tau)= {\mathcal O}(k^2 c_{st}^2 \tau^2)$ which is always smaller than $1$ when the corresponding wavelengths are larger than the sound horizon at the corresponding epoch.

The conclusions reached so far do not assume any specific background evolution.  Thus the illustrative example of a radiation-dominated plasma must corroborate the general results of Eqs. (\ref{NINETEEN})--(\ref{TWENTYONE}) and (\ref{TWENTYTWO})--(\ref{TWENTYTHREE}). Since the total sound speed of a radiation plasma is constant (i.e. $c_{st}^2 = w = 1/3$), the mode functions of Eq. (\ref{ONEa}) can be computed in a closed form:
\begin{equation}
{\mathcal R}_{\vec{q}}(\tau) = \overline{{\mathcal R}}(\vec{q}\,) \, j_{0}(q \, c_{st}\, \tau), \qquad {\mathcal R}^{\prime}_{\vec{q}}(\tau) = - q\, c_{st} \,\overline{{\mathcal R}}(\vec{q}\,) \,j_{1}(q \, c_{st} \,\tau),
\label{22}
\end{equation}
where $j_{0}(q \, c_{st} \,\tau)$ and $j_{1}(q \, c_{st} \,\tau)$ are spherical Bessel functions of 
zeroth- and first-order. To identify more easily the various different contributions in the effective anisotropic stress the sound speed  has been kept constant but generic in Eq. (\ref{22}) (we shall eventually set $c_{st} \to 1/\sqrt{3}$ only at the very end). Since ${\mathcal R}(\vec{q}\,)$ is a scalar random field, 
its correlation function and the associated power spectrum is:
\begin{equation}
\langle \overline{{\mathcal R}}(\vec{q}\,) \, \overline{{\mathcal R}}(\vec{q}^{\,\,\prime}) \rangle = \frac{2 \pi^2}{q^3} \overline{P}_{{\mathcal R}}(q) \, \delta^{(3)}(\vec{q} + 
\vec{q}^{\,\,\prime}), \qquad \overline{P}_{{\mathcal R}}(q) = {\mathcal A}_{{\mathcal R}} \biggl(\frac{q}{q_{p}}\biggr)^{n_{s} -1},
\label{22a}
\end{equation}
where ${\mathcal A}_{{\mathcal R}}$ is the amplitude of the power spectrum at the pivot scale $q_{p} = 0.002\,\mathrm{Mpc}^{-1}$ corresponding to a frequency $\nu_{p} = 2 \pi q_{p} = 3\times10^{-18}$ Hz; $0.9 < n_{s} < 1$ denotes the scalar spectral index. With the same notation employed in Eq. (\ref{22a}) the tensor power spectrum is:
\begin{equation}
\langle \overline{h}_{i\,j}(\vec{q}\,) \,\overline{h}_{m\, n} (\vec{q}^{\,\,\prime}) \rangle = \frac{2 \pi^2}{q^3} \, {\mathcal S}_{i\,j\,m\,n}(\hat{q}) \,\overline{P}_{T}(q) \, \delta^{(3)}(\vec{q} + \vec{q}^{\,\,\prime}), \qquad \overline{P}_{T}(q) = {\mathcal A}_{T}  \biggl(\frac{q}{q_{p}}\biggr)^{n_{T}},
\label{22b}
\end{equation}
where ${\mathcal S}_{i\, j\, m\, n}(\hat{q}) = [ p_{im}(\hat{q}) p_{jn}(\hat{q}) + p_{in}(\hat{q}) p_{jm}(\hat{q}) - p_{ij}(\hat{q}) p_{mn}(\hat{q})]/4$ and $p_{ij}(\hat{q}) =[ \delta_{ij} - \hat{q}_{i} \hat{q}_{j}]$. According to
the standard notations, ${\mathcal A}_{T} = r_{T} \, {\mathcal A}_{{\mathcal R}}$ 
is the amplitude of the tensor power spectrum at the same pivot scale used for the scalars.
The tensor to scalar ratio $r_{T}$ and the spectral index $n_{T}$ may be related by the so-called consistency relations (i.e. $n_{T} \simeq r_{T}/8$) but this point is not central for the present discussion.
In terms of the tensor random fields entering Eq. (\ref{22b}) the homogeneous solution of Eq. (\ref{SEVEN}) is
\begin{equation}
\overline{h}_{i\,j}(\vec{q},\tau) = \overline{h}_{i\,j}(\vec{q}\,) \, j_{0}(q\,\tau), \qquad H_{ij}(\vec{q},\tau) = \partial_{\tau} \overline{h}_{i\,j}(\vec{q},\tau) = - q \,\, \overline{h}_{i\,j}(\vec{q}\,) \, j_{1}(q\,\tau).
\label{22c}
\end{equation}
When Eqs. (\ref{22}) and (\ref{22a}) are inserted into Eqs. (\ref{TEN}) and (\ref{FOURTEEN})
 the effective anisotropic stresses obey the following concise expression:
 \begin{equation}
\Pi_{ij}^{(X)}(q,\tau) = - \frac{1}{(2\pi)^{3/2} \, \ell_{P}^2 \, a^2} \int d^{3} k\,\, k_{i} \, k_{j} \,\, \overline{{\mathcal R}}(\vec{k}) \, 
\overline{{\mathcal R}}(\vec{q} -\vec{k}) \, M^{(X)}(k \, c_{st}\,\tau, \, |\vec{q} - \vec{k}| \, c_{st} \, \tau),
\label{23}
\end{equation}
which is actually more general than the examples we are now describing. A relevant property 
of $M^{(X)}(z, w)$ is that it is generically symmetric for $w\to z$ and $z\to w$. 
In the particular case of the radiation-dominated plasma the exact expressions of $M^{(X)}(z,\,w)$ (for $X = L,\, U$) 
are\footnote{To avoid possible confusions we mention that the variable $z$ appearing in Eqs. (\ref{24}) and (\ref{25}) 
has nothing to do with $z_{t}$ introduced in Eq. (\ref{ONEa}). }
\begin{eqnarray}
&& M^{(L)}(z,w) = 4 \biggl[ j_{0}(z) j_{0}(w) - 6 c_{st}\biggl(  \frac{j_{0}(z)j_{1}(w)}{w}  +\frac{  j_{1}(z) j_{0}(w)}{z} \biggr)
- \frac{54 c_{st}^2}{w \, z} j_{1}(w) j_{1}(z)\biggr],
\label{24}\\
&& M^{(U)}(z,w) = 6 c_{st}^2 \, \biggl(\frac{z}{w} + \frac{w}{z} \biggr) j_{1}(w)  j_{1}(z).
\label{25}
\end{eqnarray}

After inserting Eq. (\ref{23}) into Eq. (\ref{SEVEN}) the tensor amplitude $h_{\lambda}(\vec{q},\tau)$ follows by recalling the explicit expressions of the Green's functions  during the radiation-dominated stage i.e. $G[q(\xi- \tau)] = -\sin{[q (\xi -\tau)]}/[a(\tau)\, q]$ and $\widetilde{\,G\,}[q(\xi - \tau)] = a(\xi) \cos{[q (\xi -\tau)]}/a(\tau)$. The energy density of the relic gravitons is then determined by plugging the results of Eq. (\ref{SEVEN})  into Eq. (\ref{FIFTEEN}). This procedure has been already discussed for an analog situation of waterfall fields in Ref. \cite{ANIS18}; thus the spectral energy density of the relic gravitons in critical units is:
\begin{eqnarray}
\Omega^{(X)}_{gw}(q,\tau) &=&  \frac{ q^2 \overline{P}_{T}(q)}{24\, H^2 \, a^2 \, |q\tau|^2} \biggl[ 1 + \frac{\sin{q\tau}}{q^2\tau^2} - \frac{\sin{2 q\tau}}{q\tau} \biggr]
+ \frac{q^3}{12}  \biggl(\frac{a_{1}^4 \, H_{1}^2}{a^4 \, H^2} \biggr)  \int_{-1}^{1} \, d\mu\, (1- \mu^2)^2
\nonumber\\
&\times&\int d k \, k^6\, \frac{\overline{P}_{\mathcal R}(k) \,\, \overline{P}_{\mathcal R}(|\vec{q} - \vec{k}|)}{k^3\,\,|\vec{q} - \vec{k}|^3}\, \biggl[ \overline{I}^{(X)\,2 }(\vec{k},\, \vec{q},\, \tau)+ \overline{J}^{(X)\,2 }(\vec{k},\, \vec{q},\, \tau) \biggr],
\label{30}
\end{eqnarray}
where $\overline{I}^{(X)}(\vec{k},\, \vec{q},\, \tau)$ and $\overline{J}^{(X)}(\vec{k},\, \vec{q},\, \tau)$ are given by:
\begin{eqnarray}
\overline{I}^{(X)}(\vec{k},\, \vec{q},\, \tau) &=& \int_{\tau_{i}}^{\tau} \,\xi \, \sin{[q (\xi -\tau)]}\, M^{(X)}(k\, c_{st} \xi;\, |\vec{q} - \vec{k}|\, c_{st} \xi) \, d\xi,
\nonumber\\
\overline{J}^{(X)}(\vec{k},\, \vec{q},\, \tau) &=& \int_{\tau_{i}}^{\tau} \, \xi \, \cos{[q (\xi -\tau)]}\, M^{(X)}(k\, c_{st} \xi;\, |\vec{q} - \vec{k}|\, c_{st} \xi) \, d\xi.
\label{31}
\end{eqnarray}
For an explicit evaluation of Eq. (\ref{31}) we recall that in the limits $z= k c_{st} \xi \gg 1$, $w= |\vec{q} - \vec{k}| c_{st} \xi \gg 1$ and $q\, c_{st} \xi \gg 1$, Eqs. (\ref{24}) and (\ref{25}) become:
\begin{equation}
M^{(L)}(z,\, w) \to \frac{ 4 \sin{z} \, \sin{w}}{w \, z} + .\,.\,.\,,\qquad M^{(U)}(z,\, w) \to \frac{ 12 c_{st}^2 \cos{z} \, \cos{w}}{w \, z}+ .\,.\,.\,.
\label{32b}
\end{equation}
Equations (\ref{32b}) apply when the wavelengths are all inside sound horizon (i.e. $k\, c_{st}/(a H) > 1$); however since $c_{st} \leq 1$ (and $k/(a\, H)> c_{st}^{-1}$) the wavelenghts are also inside the Hubble radius (i.e. $k/(a\, H)> 1$).  
The spectral energy density of the relic gravitons  inside the Hubble radius in its full form 
(i.e. including the second-order corrections) follows from Eqs. (\ref{30}) and (\ref{31}). The results of Eq. (\ref{32b}) coincide up o a phase (because  $c_{st}= 1/\sqrt{3}$). Thus the expressions of Eq. (\ref{31}) (for $X = L$ and $X =U$) will eventually inherit a phase difference that however disappears after squaring and summing up the contributions of the two integrals in each case. The common value of spectral energy density inside the 
sound horizon  is therefore 
\begin{eqnarray}
\Omega^{(U)}_{gw}(q,\tau_{0})= \Omega^{(L)}_{gw}(q,\tau_{0}) &=& \frac{r_{T} \, {\mathcal A}_{{\mathcal R}}\Omega_{R0} }{12}   \biggl(\frac{q}{q_{p}} \biggr)^{n_{T}}\biggl[ 1 + \frac{96 \, \pi^2 {\mathcal A}_{\mathcal R}}{5 r_{T}} f(n_{s}, q) \biggl(\frac{q}{q_{p}}\biggr)^{ 2 (n_{s} -1) - n_{T}} \biggr],
\label{33}\\
f(n_{s}, q) &=& a_{1}(n_{s}) + a_{2}(n_{s})  \biggl(\frac{q_{p}}{q} \biggr)^{n_{s} +1} + 
a_{3}(n_{s}) \biggl(\frac{q_{max}}{q} \biggr)^{2n_{s} -5},
\label{34}
\end{eqnarray}
where $\tau_{0}$ denotes the present value of the conformal time coordinate while $a_{i}(n_{s})$ (with 
$i=1,\,2\,3$) are three numerical constants\footnote{Even if the explicit expressions are immaterial for the 
present discussion we have that $a_{1}(n_{s}) = (n_{s} -6)/[(2 n_{s} -5) (n_{s}+1)]$, $ a_{2}(n_{s}) =  - 1/(n_{s} +1)$
and  $a_{3}(n_{s}) = 1/(2 n_{s} -5)$.}. The expressions of the coefficients $a_{i}(n_{s})$ follow from the integration of Eq. (\ref{30}) 
first over $\mu$ and then over $k$ between $q_{p}$ and $q_{max}$. The integration over $k$ can be approximated in two 
separate regions (i.e. $k<\,q$ and $k>\,q$); this way of approximating the integrals compares quite well with the 
numerical results as explicitly discussed in the case of waterfall fields \cite{ANIS18} where the power spectra appearing in the convolutions  have larger slopes but similar analytical expressions. Since $\nu_{p} = 2 \pi q_{p}$ is in the aHz region (see discussion after Eq. (\ref{22a})) 
 and $\nu_{max} = 2\pi q_{max} =190$ MHz we have that $f(n_s,\,q)= {\mathcal O}(10^{-2})$  for typical scalar 
 spectral indices $0.9< n_{s}< 1$ .

Let us finally consider Eqs. (\ref{30}) and (\ref{31}) when the corresponding wavelengths are outside the sound horizon. In this limit the asymptotic forms of Eqs. (\ref{24}) and (\ref{25}) are $M^{(L)}(z, w) \to 4(6 c_{st}^2 - 4 c_{st} +1)$ and $M^{(U)}(z, w) \to 2 c_{st}^2 [ z^2 + w^2]/3$
respectively. Once again, with the help of these asymptotic expressions 
the integrals $\overline{I}^{(X)}(\vec{k},\, \vec{q},\, \tau)$ and $\overline{J}^{(X)}(\vec{k},\, \vec{q},\, \tau)$ of Eq. (\ref{31}) can be estimated. The first-order contribution has the standard form valid during the radiation-dominated phase and it follows from the first term at the right-hand side of Eq. (\ref{30}) for $q\tau \ll 1$;
the second-order correction is however different in the two gauges so that the general form of $\Omega_{gw}^{(X)}(q,\tau)$ is:
 \begin{equation}
\Omega^{(X)}_{gw}(q,\tau) =\overline{\Omega}_{gw}(q,\tau)\biggl[ 1 + \omega^{(X)}_{gw}(q, \tau) \biggr],\qquad \overline{\Omega}_{gw}(q,\tau) =  \frac{r_{T}{\mathcal A}_{{\mathcal R}}}{12} q^2 \tau^2 \biggl(\frac{q}{q_{p}}\biggr)^{n_{T}},
 \label{36}
 \end{equation}
 where the two functions $\omega^{(L)}(q,\tau_{0})$ and $\omega^{(U)}(q,\tau_{0})$ are: 
 \begin{eqnarray}
\omega^{(L)}_{gw}(q,\tau) &=& \frac{64}{15} \frac{{\mathcal A}_{\mathcal R}}{r_{T}} \, \Omega_{R0} \, q^2 \tau^2 \, \biggl[ 1 + \frac{q^2 \tau^2}{9} \biggr]\, \biggl(\frac{q}{q_{p}}\biggr)^{2(n_{s} -1) -n_{T}} \overline{f}^{(L)}(n_{s},\, q),
\nonumber\\
 \omega^{(U)}_{gw}(q,\tau) &=& \frac{4}{135} \frac{{\mathcal A}_{\mathcal R}}{r_{T}} \, \Omega_{R0} \, q^6 \tau^6 \, \biggl[ 1 + \frac{q^2 \tau^2}{25} \biggr]\, \biggl(\frac{q}{q_{p}}\biggr)^{2(n_{s} -1)-n_{T}} \overline{f}^{(U)}(n_{s},\, q). 
 \label{37}
\end{eqnarray}
The form of  $\overline{f}^{(L)}(n_{s},\, q)$ and $\overline{f}^{(U)}(n_{s},\, q)$ is not central to the present discussion and it is anyway similar to $f(n_{s}, q)$ appearing in Eq. (\ref{34}). What matters here
is the parametric dependence of the correction upon $q\,\tau$, i.e. $ \omega^{(U)}_{gw}(q,\tau)/\omega_{gw}^{(L)}(q,\tau) =  {\mathcal O}(|q\tau|^4)$. This result coincides with the general conclusion of 
 Eq. (\ref{TWENTYTHREE}): there we considered the anisotropic stress itself while Eqs. (\ref{36}) and (\ref{37}) 
are quadratic in the anisotropic stresses. This is why the mismatch between the two expressions 
is not given by $|q\tau|^2$ (as in Eq. (\ref{TWENTYTHREE})) but by the square of it. 

The effective anisotropic stresses computed in different coordinate systems 
coincide inside the sound horizon while they diverge in the opposite limit. 
 The present approach does not assume any specific background evolution 
but it has been explicitly corroborated by  the analysis of a radiation 
dominated plasma. The same kind of spurious gauge-invariance examined here
is also manifest when the energy density of the relic gravitons is derived from competing 
energy-momentum pseudo-tensors. To lowest order the ambiguity can 
be solved (or alleviated) by selecting an energy-momentum pseudo-tensor with reasonable 
physical properties such as the one obtained long ago by Ford and Parker. The present considerations 
show that some ambiguities are likely to reappear from the higher-order processes as 
a direct consequence of the lack of localization of the energy-momentum of the gravitational field. 

The author wishes to thank T. Basaglia, A. Gentil-Beccot, S. Rohr and J. Vigen of the CERN 
Scientific Information Service for their precious collaboration.

\end{document}